\documentclass[aps,prb,twocolumn,superscriptaddress]{revtex4-2}

\usepackage{amsmath, amssymb}
\usepackage{graphicx}
\usepackage{dcolumn}
\usepackage{booktabs}
\usepackage{xcolor} 
\usepackage{soul}
\usepackage{float}
\usepackage[colorlinks = true, urlcolor = cyan, linkcolor=blue, citecolor=blue, filecolor=magenta]{hyperref}
\usepackage{adjustbox}

\newcommand{\ket}[1]{|#1\rangle}

\begin{document}

\title{Two-Level System Spectroscopy from Correlated Multilevel Relaxation in Superconducting Qubits}

\author{Tanay Roy}
\email{roytanay@fnal.gov}
\affiliation{Superconducting Quantum Materials and Systems (SQMS) Center, Fermi National Accelerator Laboratory, Batavia, IL 60510, USA}

\author{Xinyuan You}
\email{xinyuan@fnal.gov}
\affiliation{Superconducting Quantum Materials and Systems (SQMS) Center, Fermi National Accelerator Laboratory, Batavia, IL 60510, USA}

\author{David van Zanten}
\affiliation{Superconducting Quantum Materials and Systems (SQMS) Center, Fermi National Accelerator Laboratory, Batavia, IL 60510, USA}

\author{Francesco Crisa}
\affiliation{Superconducting Quantum Materials and Systems (SQMS) Center, Fermi National Accelerator Laboratory, Batavia, IL 60510, USA}

\author{Sabrina Garattoni}
\affiliation{Superconducting Quantum Materials and Systems (SQMS) Center, Fermi National Accelerator Laboratory, Batavia, IL 60510, USA}

\author{Shaojiang Zhu}
\affiliation{Superconducting Quantum Materials and Systems (SQMS) Center, Fermi National Accelerator Laboratory, Batavia, IL 60510, USA}

\author{Anna Grassellino}
\affiliation{Superconducting Quantum Materials and Systems (SQMS) Center, Fermi National Accelerator Laboratory, Batavia, IL 60510, USA}

\author{Alexander Romanenko}
\affiliation{Superconducting Quantum Materials and Systems (SQMS) Center, Fermi National Accelerator Laboratory, Batavia, IL 60510, USA}
\email{}

\date{\today}

\begin{abstract}
Transmon qubits are a cornerstone of modern superconducting quantum computing platforms. Temporal fluctuations of energy relaxation in these qubits are widely attributed to microscopic two-level systems (TLSs) in device dielectrics and interfaces, yet isolating individual defects typically relies on tuning the qubit or the TLS into resonance. We demonstrate a novel spectroscopy method for fixed-frequency transmons based on multilevel relaxation: repeated preparation of the second excited state and simultaneous $T_1$ extraction of the first and second excited states reveals characteristic correlations in the decay rates of adjacent transitions. From these correlations we identify one or more dominant TLSs and reconstruct their frequency drift over time. Remarkably, we find that TLSs detuned by $\gtrsim 100\,\mathrm{MHz}$ from the qubit transition can still significantly influence relaxation. The proposed method provides a powerful tool for TLS spectroscopy without the need to tune the transmon frequency, either via a flux-tunable inductor or AC-Stark shifts.

\end{abstract}

\maketitle

\textit{Introduction---}
Quantum information processing promises computational advantages for tasks ranging from quantum simulation to optimization and cryptography. Among the leading hardware platforms, superconducting circuits combine strong nonlinearity, fast control, and straightforward integration with microwave technology, enabling multi-qubit processors with high-fidelity gates and readout~\cite{Devoret2013,Kjaergaard2020}. A central challenge is the preservation of quantum coherence. In state-of-the-art transmons, energy relaxation and dephasing are often limited by parasitic loss associated with microscopic two-level systems (TLSs) in amorphous dielectrics and at material interfaces~\cite{Martinis2005,Muller2019,Anderson1972,Phillips1987,Gao2008}. These defects can both reduce the average lifetime and generate temporal fluctuations in coherence, impacting calibration stability and processor performance~\cite{Burnett2014,Burnett2019Benchmark,Klimov2018Fluct,Bal2023systematic,Valieres2025LossTangent}. Despite extensive progress, the microscopic origin and dynamics of the relevant TLS defects remain under active study~\cite{Muller2019,Phillips1987,Shnirman2005Noise,You2021PNNoise}.

A direct way to detect and characterize individual TLSs is to bring a qubit transition into resonance with the defect. Early experiments observed avoided level crossings and coherence signatures of strongly coupled TLSs in Josephson devices~\cite{Simmonds2004}. More broadly, flux-tunable qubits enable the qubit frequency to be swept across a band where TLS resonances appear as sharp spectroscopic features and as localized reductions in $T_1$ when the qubit is tuned near resonance~\cite{Martinis2005,Lisenfeld2010,Shalibo2010}. Complementary approaches tune the TLS itself, for example by applying controlled strain or a DC electric field~\cite{Grabovskij2012,Lisenfeld2015,Sarabi2016Dipole}. Additional protocols resolve TLS parameters via decoherence or noise spectroscopy of individual defects~\cite{Lisenfeld2016SciRep,Matityahu2017RabiNoise}. Such tunability-based TLS spectroscopy provides access to defect frequencies, coupling strengths, and frequency drift over time~\cite{Burnett2014,Burnett2019Benchmark,Klimov2018Fluct}, and it provides key experimental support for microscopic mechanisms behind coherence fluctuations.

Fixed-frequency transmons are nevertheless widely used in scalable architectures because they avoid added control complexity and reduce sensitivity to low-frequency flux noise~\cite{IBM2024two_qubit_gate}. However, the inability to sweep the qubit frequency complicates direct TLS identification and motivates diagnostics that do not rely on flux tuning (or large AC-Stark shifts that may be impractical), while retaining sensitivity to individual, off-resonant defects. This is particularly important because TLSs can affect relaxation even when appreciably detuned from the qubit transition of interest, and because different transitions of a weakly anharmonic transmon probe the noise spectrum at distinct frequencies. In addition, actively engineering the TLS noise spectrum provides a potential route to stabilize and improve coherence~\cite{You2022NoiseEngineering}.

Here we present a simple yet powerful TLS-characterization method tailored to fixed-frequency transmons, based on monitoring multilevel relaxation. By repeatedly preparing the second excited state and extracting the time-dependent lifetimes of the first and second excited states, we observe correlated fluctuations of decay rates for several days. We show that these correlations provide a sensitive fingerprint of one (or a few) dominant TLSs and enable reconstruction of their frequency fluctuations over time, while directly demonstrating that TLSs detuned by more than $100\,\mathrm{MHz}$ from the qubit transition can still significantly limit coherence.

\begin{figure}
    \centering
    \includegraphics[width=\columnwidth]{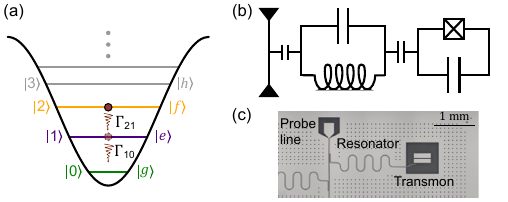}
    \caption{The transmon device (a) The anharmonic energy eigenstates of a transmon circuit. The $\ket{2} \rightarrow \ket{1}$ and $\ket{1} \rightarrow \ket{0}$ decay rates $\Gamma_{21}$ and $\Gamma_{10}$ can significantly deviate from the bosonic relation $\Gamma_{21} = 2\Gamma_{10}$. (b) A schematic of the electrical circuit showing the transmon (right) coupled to a readout resonator (middle) and probed through a transmission line (left). (c) Optical image the device.}
    \label{fig:fig1}
\end{figure}

\textit{Protocol---} We consider the first three energy levels of a transmon circuit labeled as $\ket{0}, \ \ket{1}$ and $\ket{2}$ (see Fig.~\ref{fig:fig1}(a)). The transition frequencies $\omega_{01}$ and $\omega_{12}$ between the $\ket{0} \leftrightarrow \ket{1}$  and $\ket{1} \leftrightarrow \ket{2}$ levels, respectively, differ by anharmonicity $\alpha<0$ so that $\omega_{12} = \omega_{01} + \alpha$. The populations ($P_n, n\in \{0,1,2 \}$) of the three levels satisfy the following equations:
\begin{equation}
\begin{aligned}
    P_2'(t) &= -\Gamma_{21} P_2(t), \\
    P_1'(t) &= -\Gamma_{10} P_1(t) + \Gamma_{21} P_2(t), \\
    P_0'(t) &= +\Gamma_{10} P_1(t),
\label{eq:diff_eqn}
\end{aligned}
\end{equation}
where $\Gamma_{21}$ and $\Gamma_{10}$ are, respectively, the decay rates of $\ket{2} \rightarrow \ket{1}$ and $\ket{1} \rightarrow \ket{0}$ processes due to natural relaxation. Here $P'_n(t)$ denotes the time derivative.

When the transmon is excited to $\ket{2}$, we use the initial conditions $P_2(0)=1, P_1(0)=P_0(0)=0$ to obtain the following equations of motion:
\begin{equation}
\begin{aligned}
    P_2(t) &= e^{-\Gamma_{21} t}, \\
    P_1(t) &= \dfrac{e^{-\Gamma_{10} t} - e^{-\Gamma_{21} t}}{\Gamma_{21} - \Gamma_{10}}, \\
    P_0(t) &= \dfrac{(1-e^{-\Gamma_{10} t})\Gamma_{21} - (1-e^{-\Gamma_{21} t})\Gamma_{10}}{\Gamma_{21} - \Gamma_{10}}.
\label{eq:P_n}
\end{aligned}
\end{equation}
Experimentally measuring the populations of the three levels as a function of time thus enables simultaneous extraction of relaxation times of the first ($\ket{1}\equiv\ket{e}$) and second ($\ket{2}\equiv\ket{f}$) excited states defined as $T_\text{1e}=1/\Gamma_{10}$ and $T_\text{1f}=1/\Gamma_{21}$, respectively. Note that, for simplicity, we have not included the effect of heating as its effect is usually negligible for devices operating at dilution temperatures~\cite{Egger2018thermal, Heinsoo2018thermal, Kulikov2020thermal, kim2025two_cell, Jin2015thermal_3D}.

\begin{figure}
    \centering
    \includegraphics[width=\columnwidth]{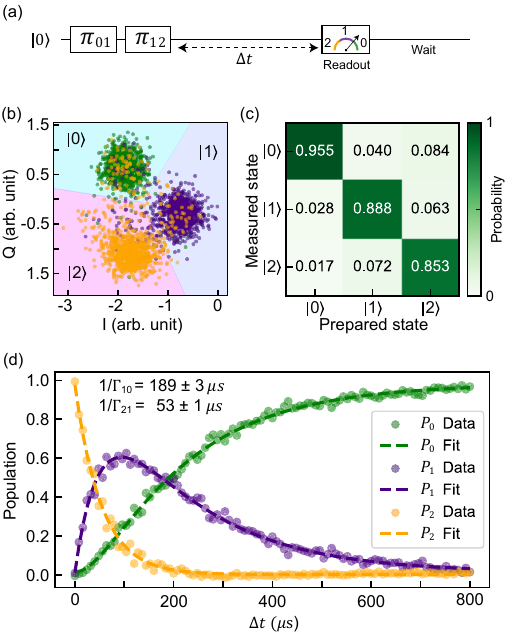}
    \caption{$T_1$ experiment with three-levels. (a) Pulse sequence for the $T_1$ experiment. The two $\pi$ pulses acting on the ground state $\ket{0}$ bring the transmon to the second excited state $\ket{2}$. A three-level readout is performed after a variable period $\Delta t$. A waiting time ensures the transmon's relaxation to the ground state. (b) Measured I-Q blobs for the first three energy eigenstates of the transmon. The discrimination boundaries are calculated to minimize assignment error. (c) Confusion matrix obtained from (b) resulting in an assignment fidelity of $\mathcal{F}_3=89.9\%$. (d) Populations traces from a typical $T_1$ experiment. The decay rates are extracted by simultaneously fitting populations of each levels as a function of $\Delta t$.}
    \label{fig:fig2}
\end{figure}

\textit{Experimental Results---} We perform experiments with several uncoupled 2D transmon devices (see Fig.~\ref{fig:fig1}(b)) made using Tantalum-capped Niobium as base metal and Al/AlOx Josephson junction on HEMEX-grade sapphire substrates~\cite{Bal2023systematic}. The transmons are dispersively coupled to individual $\lambda/4$ resonators which are measured in a hangar geometry as shown in Fig.~\ref{fig:fig1}(c). The devices are cooled down to about 8 mK inside a dilution refrigerator. All devices are measured using the RFSoC ZCU216 board. The experimentally parameters of the two representative transmons are provided in Table~\ref{tab:params}.

\begin{table*}[t]
\caption{Parameters for the two transmon devices.}
\label{tab:params}
\centering
\begin{ruledtabular}
\begin{tabular}{ccccccc}
Device
& $\omega_{01}/2\pi$ (MHz)
& $\alpha/2\pi$ (MHz)
& Median $T_\text{1e}$ ($\mu$s)
& Median $T_\text{1f}$ ($\mu$s)
& $\omega_{\mathrm{rdt}}/2\pi$ (MHz)
& Readout fidelity $\mathcal{F}_3$ \\
\hline
A & 4822.08 & $-280.37$ & 155 & 64 & 7049.05 & 89.9\% \\
B & 5810.32 & $-201.32$ & 111 & 90 & 7714.24 & 90.1\% \\
\end{tabular}
\end{ruledtabular}
\end{table*}

\begin{figure*}
    \centering
    \includegraphics[width=0.9\textwidth]{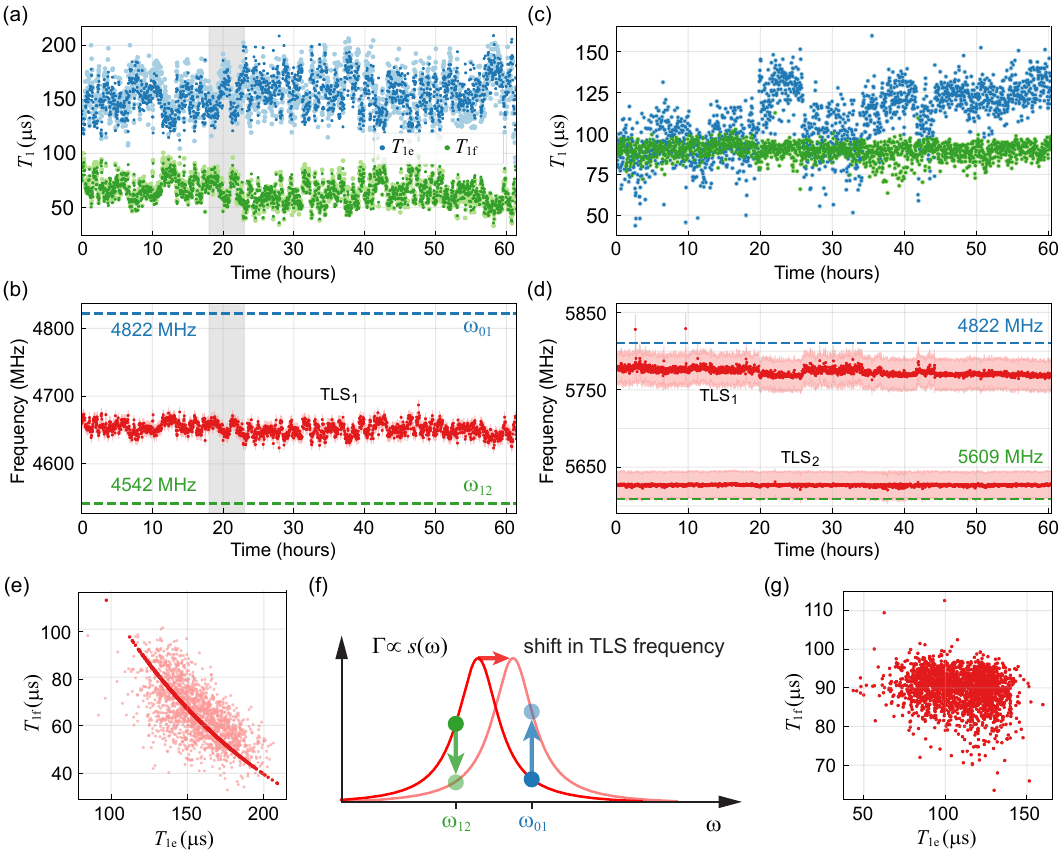}
    \caption{
    Synchronous measurement of transmon relaxation times between different levels. (a) Measured relaxation times $T_\text{1e}$ (transparent blue) and $T_\text{1f}$ (transparent green) of device A as a function of time. A representative region of strong anti-correlation is highlighted by the gray bar. Solid dots denote fits to a single–TLS model.  
    (b) TLS frequency extracted from the single-TLS fits in (a) as a function of time; the shaded bar indicates the extracted TLS linewidth. Dashed lines show the transmon transition frequencies $\omega_{01}$ (blue) and $\omega_{12}$ (green) for reference.  
    (c-d) Same measurements and analysis for device B, where $T_\text{1e}$ and $T_\text{1f}$ exhibit minimal correlation, requiring a two-TLS model for accurate fitting. 
    (e) Correlation plot of measured $T_\text{1e}$ and $T_\text{1f}$ for device A. Solid and transparent dots represent fitted and measured data, respectively. (f) Schematics illustrating the change in qubit transition rates due to the a TLS frequency shift. (g) Correlation plot for device \textit{B}.}
    \label{fig:fig3}
\end{figure*}

To obtain the decay rates, $T_1$ experiments are performed after initializing the transmon in the $\ket{2}$ state and monitoring the population as a function of times. The preparation of $\ket{1}$ requires a $\pi_{10}$ pulse starting from the ground state, whereas a following $\pi_{21}$ pulse prepares $\ket{2}$ as illustrated in Fig.~\ref{fig:fig2}(a). In order to identify the three states, an initial readout optimization is performed where measured readout signals are first plotted on the complex plane. A 3-way classification is performed to obtain the highest assignment fidelity by optimizing the demarcation boundaries. Then the frequency, amplitude and duration of the readout pulse are optimized for achieving the lowest readout error. The measured IQ blobs for device~\textit{A} are shown in Fig.~\ref{fig:fig2}(b). The resulting confusion matrix is presented in Fig.~\ref{fig:fig2}(c) showing a 3-level readout fidelity $\mathcal{F}_3=0.899$. The inverted confusion matrix is multiplied to the raw population vectors at each time point to correct for the imperfect qutrit readout~\cite{Roy2023qutrit}. Typical curves after correcting for readout error are plotted in Fig.~\ref{fig:fig2}(d). The curves are simultaneously fitted to Eq.~\eqref{eq:P_n} to obtain $T_\text{1e}$ and $T_\text{1f}$.


Figure~\ref{fig:fig3} shows the measured transmon relaxation times between different energy levels. For device~\textit{A}, the measured values of $T_\text{1e}$ and $T_\text{1f}$ over a period exceeding 60 hours are plotted as transparent dots in Fig.~\ref{fig:fig3}(a). On average, the lifetime of $T_\text{1f}$ is about half of that of $T_\text{1e}$, due to the increased matrix element for the higher state. Over time, both relaxation times fluctuate and show strong anti-correlation; a representative region is highlighted in gray. To visualize the anti-correlation between the measured relaxation times, we plot $(T_\text{1e},T_\text{1f})$ in Fig.~\ref{fig:fig3}(e) as transparent dots.  
The skewed distribution of points with a negative slope—rather than an isotropic cloud—indicates strong anti-correlation.  

In the literature, fluctuation of the qubit lifetime (i.e., $T_\text{1e}$) is typically attributed to variation of an individual TLS frequency near the qubit frequency (i.e., $\omega_\text{01}$). The observed anti-correlation can be explained by the same mechanism. Consider a single TLS whose frequency lies between the qubit $\ket{0}\leftrightarrow\ket{1}$ and $\ket{1}\leftrightarrow\ket{2}$ transitions.  
The corresponding noise spectral density $S(\omega)$ is a Lorentzian centered at the TLS frequency $\omega_\text{TLS}$ with linewidth $\gamma_\text{TLS}$.  
This results in decay of both the $|2\rangle$ and $|1\rangle$ levels, with decay rates proportional to $S(\omega_\text{12})$ and $S(\omega_\text{01})$, respectively.  
In the presence of TLS frequency fluctuation, e.g., when $\omega_\text{TLS}$ shifts toward $\omega_\text{01}$ and away from $\omega_\text{12}$, $S(\omega_\text{01})$ increases while $S(\omega_\text{12})$ decreases, leading to the observed anti-correlation of relaxation times for the two levels. A schematics illustrating the above mechanism is shown in Fig.~\ref{fig:fig3}(f).  

To quantitatively analyze the data, we fit using the single-TLS model, with the TLS frequency $\omega_\text{TLS}(t)$, its linewidth $\gamma_\text{TLS}$, and a scaling factor determined by the coupling strength (see Appendix A).  
The fitted results are shown in Fig.~\ref{fig:fig3}(a) as solid dots.  
Although the model is simple, it captures the observed anti-correlation very well. The extracted TLS frequency as a function of time is shown in Fig.~\ref{fig:fig3}(b), with the transition frequencies $\omega_\text{01}$ and $\omega_\text{12}$ indicated by dashed lines.  
In Fig.~\ref{fig:fig3}(e), we plot the correlation using the fitted data as solid dots. The resulting curve with perfect anti-correlation, reproduces the main trend of the measured points. The spread of the experimental data could be due to other noise sources, such as the presence of additional weakly coupled TLSs. 

We perform the same measurement for another device \textit{B}, with results shown in Fig.~\ref{fig:fig3}(c). Unlike the previous case, here $T_\text{1e}$ and $T_\text{1f}$ show minimal anti-correlation: $T_\text{1e}$ exhibits strong fluctuation over time, while $T_\text{1f}$ remains stable. Notably, there are intervals where the lifetime of the higher level exceeds that of the lower level. Both features point to the presence of more than a single TLS. Indeed, the data can be fit with a two-TLS model. The extracted frequencies of the two TLSs are shown in Fig.~\ref{fig:fig3}(d). TLS$_1$ lies closer to the $\ket{0}\leftrightarrow\ket{1}$ transition and fluctuates strongly over time, explaining the variations of $T_\text{1e}$, while TLS$_2$ remains stable during the measurement, consistent with the nearly constant $T_\text{1f}$.  
For visualization, we plot the correlation for device~\textit{B} in Fig.~\ref{fig:fig3}(g). In sharp contrast with Fig.~\ref{fig:fig3}(e), here the data are more isotropic, indicating weak correlation between $T_\text{1e}$ and $T_\text{1f}$.

\textit{Conclusion and discussion---} We introduced a multilevel-relaxation protocol for identifying and tracking individual TLS defects in \emph{fixed-frequency} transmons by monitoring the time-dependent lifetimes of multiple transitions. Repeated preparation of $\ket{2}$ followed by a three-level $T_1$ analysis yields simultaneous estimates of $T_{\rm 1e}(t)$ and $T_{\rm 1f}(t)$, providing a dissipation fingerprint that is more diagnostic than monitoring a single transition alone.

For device~\textit{A}, we observe pronounced anti-correlation between $T_{\rm 1e}$ and $T_{\rm 1f}$ over a measurement window exceeding 60 hours. This behavior is quantitatively captured by a simple single-TLS model in which one dominant fluctuator lies between $\omega_{01}$ and $\omega_{12}$, so that a drift of $\omega_{\rm TLS}(t)$ simultaneously increases loss at one transition while decreasing it at the other. The resulting fits enable reconstruction of the TLS frequency trajectory and show frequency wander on the order of $\sim 10\,\mathrm{MHz}$, much larger than typical fluctuations of the transmon transition frequencies ($\lesssim 10\,\mathrm{kHz}$)~\cite{abdisatarov2025mag_studies, Burnett2014}. In contrast, device~\textit{B} exhibits weak correlation and intermittent inversion $T_{\rm 1e}<T_{\rm 1f}$, pointing to multiple relevant defects and motivating a two-TLS description.

A key practical implication is that TLS-induced loss cannot be assessed solely by proximity to the qubit transition. Even a defect detuned by $\gtrsim 100\,\mathrm{MHz}$ from $\omega_{01}$ can measurably limit relaxation through the Lorentzian tails of its noise spectral density sampled at nearby transition frequencies. More broadly, the relative behavior of $\Gamma_{21}$ and $\Gamma_{10}$ directly quantifies deviations from the bosonic expectation $\Gamma_{21}=2\Gamma_{10}$ and thus provides a sensitive probe of frequency-selective loss channels in weakly anharmonic circuits.

The method is intentionally minimal in experimental overhead, but its interpretation relies on a few assumptions. We treat $\omega_{01}$ and $\omega_{12}$ as constant and attribute the dominant temporal variation to drifting TLS frequencies; this is well motivated by the separation of scales noted above~\cite{abdisatarov2025mag_studies, Burnett2014}. We also neglect heating-induced upward transitions, which would add terms such as $\Gamma_{01}P_0(t)$ and $\Gamma_{12}P_1(t)$ to Eq.~\eqref{eq:diff_eqn}; these corrections are expected to be negligible at mK temperatures where modern devices can reach thermal populations $<1\%$ in $\ket{1}$ and $<0.01\%$ in $\ket{2}$~\cite{Egger2018thermal, Heinsoo2018thermal, Kulikov2020thermal, kim2025two_cell, Jin2015thermal_3D}. In future work, extending the protocol to additional levels (e.g., including $\ket{3}$~\cite{Liu2023SUD4, Seifart2023ququart, Wang2025qudit}) and combining it with weak, controlled frequency shifts (e.g., small AC-Stark shifts) could further disambiguate multiple nearby TLSs and improve robustness in the multi-defect regime.

\begin{acknowledgments}
This work was supported by the U.S. Department of Energy, Office of Science, National Quantum Information Science Research Centers, Superconducting Quantum Materials and Systems Center (SQMS), under Contract No. 89243024CSC000002. Fermilab is operated by Fermi Forward Discovery Group, LLC under Contract No. 89243024CSC000002 with the U.S. Department of Energy, Office of Science, Office of High Energy Physics. This work made use of the Pritzker Nanofabrication Facility of the Institute for Molecular Engineering at the University of Chicago, which receives support from Soft and Hybrid Nanotechnology Experimental (SHyNE) Resource (NSF ECCS-1542205), a node of the National Science Foundation’s National Nanotechnology Coordinated Infrastructure. 
\end{acknowledgments}

\appendix

\section{TLS fitting model}

Following a standard TLS model, the noise spectral density of a single TLS is 
\begin{equation}
    s(\omega) =  A\,\frac{\gamma_\text{TLS}}{(\omega-\omega_\text{TLS})^2 + \gamma_\text{TLS}^2},
\end{equation}
where $\gamma_\text{TLS}$ and $\omega_\text{TLS}$ denote the linewidth and frequency of the TLS. The prefactor $A$ accounts for the polarization of the TLS. When multiple TLSs are coupled to the same qubit, the resulting transition rate at a given time $t$ follows 
\begin{equation}
    \Gamma_{j+1,j}(t) = \sum_n B_n\,\frac{\gamma_{\text{TLS},n}}{(\omega_{j,j+1}-\omega_{\text{TLS},n}(t))^2 + \gamma_{\text{TLS},n}^2},
\end{equation}
where $B_n$ incorporates the qubit transition matrix element and coupling to the $n$-th TLS, in addition to the TLS polarization. 
Here, $\omega_{j,j+1}$ is taken to be the $\ket{0}\leftrightarrow\ket{1}$ or $\ket{1}\leftrightarrow\ket{2}$ transition frequency (constant) in the fitting for time-dependent $T_{1,j}(t)$ with $T_{\rm 1e}(t)=1/\Gamma_{10}(t)$ and $T_{\rm 1f}(t)=1/\Gamma_{21}(t)$ measured for different levels. 
We use least-square fitting model, with fitting parameters: $B_n$, $\gamma_{\text{TLS},n}$, and time-dependent $\omega_{\text{TLS},n}(t)$ to obatin a simultanous fit of $T_{\rm 1e}(t)$ and $T_{\rm 1f}(t)$. 

\section{Measurement error mitigation}

We consider a model in which the preparation of quantum states is assumed to be perfect, and all imperfections are attributed to the measurement process. In an ideal, error-free measurement, the probability $P(j|k)$ of obtaining outcome $j$ when the basis state $k$ is prepared is 1 if $j=k$ and zero otherwise. In realistic experiments, noise and systematic errors cause deviations from this ideal behavior and we aim to characterize these measurement errors and mitigate their effect.

To perform measurement error mitigation for a single qutrit, we prepare each of the three basis states a large number of times followed by immediate measurements. From the resulting IQ blobs and 3-way classification, we construct the assignment matrix 
$\mathbf{M}_{j,k} = P(j|k)$ where $j,k \in \{0,1,2\}$. Here, the $k$-th column of $\mathbf{M}$ represents the probability distribution over measurement outcomes $j$ when the initial state $k$ is prepared. The matrix $\mathbf{M}$ is therefore a $3 \times 3$ stochastic matrix that fully characterizes the measurement errors of the qutrit as shown in the main text.

Let $\vec{P}_{\mathrm{id}}=(P_0, P_1, P_2)^\mathrm{T}$ be the ideal probability vector obtained from a projective measurement. Due to measurement errors, the experimentally observed probability vector $\vec{P}_{\mathrm{expt}}$ is related to the ideal one by
$\vec{P}_{\mathrm{expt}} = \mathbf{M}\vec{P}_{\mathrm{id}}$. Provided that $\mathbf{M}$ is invertible, the ideal probability distribution can be recovered by $\vec{P}_{\mathrm{id}} = \mathbf{M}^{-1}\vec{P}_{\mathrm{expt}}.$
The corrected probability vector $\vec{P}_{\mathrm{id}}$ is used for subsequent analysis.





\bibliography{main}

\end{document}